\documentclass[a4paper,12pt]{article}

\usepackage{amsmath}
\usepackage[psamsfonts]{amssymb}
\usepackage{rsfs}
\usepackage{bm}

\usepackage{cite}
\usepackage[dvips]{graphicx}
\usepackage{color}

\begin{document} 

\begin{titlepage}

\baselineskip 10pt
\hrule 
\vskip 5pt
\leftline{}
\leftline{Chiba Univ./KEK Preprint
          \hfill   \small \hbox{\bf CHIBA-EP-155}}
\leftline{\hfill   \small \hbox{\bf KEK Preprint 2005-61}}
\leftline{\hfill   \small \hbox{hep-lat/0509069}}
\leftline{\hfill   \small \hbox{October 2005}}
\vskip 5pt
\baselineskip 14pt
\hrule 
\vskip 1.0cm
\centerline{\Large\bf 
Lattice construction of 
} 
\vskip 0.3cm
\centerline{\Large\bf  
Cho-Faddeev-Niemi decomposition 
}
\vskip 0.3cm
\centerline{\Large\bf  
and gauge invariant monopole
}
\vskip 0.3cm
\centerline{\large\bf  
}

\vskip 0.5cm

\centerline{{\bf 
S. Kato$^{\sharp,{1}}$, 
K.-I. Kondo$^{\dagger,\ddagger,{2}}$,  
T. Murakami$^{\ddagger,{3}}$,
A. Shibata$^{\flat,{4}}$, 
T. Shinohara$^{\ddagger,{5}}$
and S. Ito$^{\star,{6}}$
}}  
\vskip 0.5cm
\centerline{\it
${}^{\sharp}$Takamatsu National College of Technology, Takamatsu 761-8058, Japan
}
\vskip 0.3cm
\centerline{\it
${}^{\dagger}$Department of Physics, Faculty of Science, 
Chiba University, Chiba 263-8522, Japan
}
\vskip 0.3cm
\centerline{\it
${}^{\ddagger}$Graduate School of Science and Technology, 
Chiba University, Chiba 263-8522, Japan
}
\vskip 0.3cm
\centerline{\it
${}^{\flat}$Computing Research Center, High Energy Accelerator Research Organization (KEK),  
}
\centerline{\it
Tsukuba 
305-0801, 
Japan
}
\vskip 0.3cm
\centerline{\it
${}^{\star}$Nagano National College of Technology, 716 Tokuma, Nagano City, 381-8550, Japan
}
\vskip 1cm

\begin{abstract}
We present the first implementation of the Cho--Faddeev--Niemi decomposition of the SU(2) Yang-Mills field on a lattice. 
Our construction retains the color symmetry (global SU(2) gauge invariance) even after a new type of Maximally Abelian gauge, as explicitly demonstrated by numerical simulations. 
Moreover, we propose a gauge-invariant definition of the magnetic monopole current using this formulation and compare the new definition with the conventional one by DeGrand and Toussaint to exhibit its validity. 

\end{abstract}


\hrule  
\vskip 0.1cm
${}^1$ 
  E-mail:  {\tt kato@takamatsu-nct.ac.jp}
  
${}^2$ 
  E-mail:  {\tt kondok@faculty.chiba-u.jp}
  
${}^3$ 
  E-mail:  {\tt tom@cuphd.nd.chiba-u.ac.jp}
  
${}^4$ 
  E-mail:  {\tt akihiro.shibata@kek.jp}
  
${}^5$ 
  E-mail:  {\tt sinohara@cuphd.nd.chiba-u.ac.jp}

${}^6$ 
  E-mail:  {\tt shoichi@ei.nagano-nct.ac.jp}

\par 
\par\noindent


\vskip 0.5cm

\newpage
\pagenumbering{roman}




\end{titlepage}


\pagenumbering{arabic}

\baselineskip 14pt
\section{Introduction}

A change of variables of the non-Abelian gauge potential in Yang--Mills theory was proposed by Cho \cite{Cho80} and Faddeev and Niemi \cite{FN98}.  The Cho-Faddeev-Niemi (CFN) decomposition or change of variables introduces a color vector field $\bm{n}(x)$ enabling us to extract explicitly  the magnetic monopole as a topological degree of freedom from the gauge potential without introducing the fundamental scalar field in Yang--Mills theory. 
The CFN decomposition has been formulated and extensively studied on the continuum spacetime. 
For non-perturbative studies, however, it is desirable to put the CFN decomposition on a lattice.  This will enable us to perform powerful numerical simulations to obtain fully non-perturbative results. 
The main purpose of this paper is to propose a lattice formulation of the CFN decomposition and to perform the numerical simulations on the lattice, paying special attention to the magnetic monopole. 

The magnetic monopole is the indispensable ingredients for the dual superconductivity scenario \cite{dualsuper} for  quark confinement. 
The idea of Abelian projection due to 't Hooft \cite{tHooft81} is that the partial gauge fixing can extract the physical degrees of freedom relevant in the long-distance of QCD.  
It has been shown that a magnetic monopole appears as a defect (singularity) of the partial gauge fixing at degenerate points of the operator to be diagonalized through the Abelian projection. 
The most efficient partial gauge fixing from this viewpoint is known to be the Maximally Abelian gauge (MAG) \cite{KLSW87}, although we can consider other Abelian gauges \cite{Sijs98}. 
The numerical simulations \cite{SY90} have confirmed that only  this gauge leads to the Abelian dominance predicted by \cite{EI82} and the magnetic monopole dominance \cite{SNW94} for the string tension.
This is also the case of chiral symmetry breaking.

In defining the monopole on the lattice, the DeGrand-Toussaint (DT) method \cite{DT80} has been extensively used so far.
However, it is not clear that the DT monopole agrees with the QCD monopole a la 't Hooft, in the sense that 
i) DT method defines a monopole by counting the number of Dirac strings coming out from an elementary cube without examining the singularities of the gauge fixing condition \cite{Sijs98},
and that 
ii) there is no correlation between the existence point of the DT monopole and the degenerate point  of the eigenvalues of the diagonalized operator  in the Abelian projection.  
Furthermore, 
iii) the monopole dominance can not be seen in Abelian gauges other than MAG, but the MAG breaks explicitly the color symmetry (global gauge invariance). This disadvantage of the MAG yields the dubious reputation that the QCD magnetic monopole might be a gauge artifact. 
Recently, some ideas of defining gauge invariant monopole on a lattice were submitted, e.g.,  \cite{GI-monopole}.  These approaches could be compared with confinement on a lattice based on other definitions \cite{Nakajima00}.

In this paper, we propose a method of implementing the CFN decomposition on a lattice. 
Our lattice formulation reflects a new viewpoint proposed by three of the authors in a previous paper \cite{KMS05}, which enables us to retain the local and global  gauge invariance even after the new type of MAG. 
Then, within this framework of the CFN decomposition, we give a definition of the gauge invariant monopole on a lattice.  The comparison of our construction with the DT one in the conventional MAG reveal that two methods give nearly the same result for the monopole density. 
 Moreover, if the lattice version of the gauge-invariant field strength $G_{\mu\nu}$ defined in \cite{KMS05} is separated into the electric and magnetic parts, the magnetic part gives the dominant contribution to the CFN monopole and the CFN monopole is determined by the hedgehog-like configuration of the color vector field $\bm{n}$ in the CFN decomposition, confirming the expectation in the previous work \cite{Kondo04,KKMSS05}.

\section{CFN decomposition in the continuum}

We adopt the Cho-Faddeev-Niemi (CFN) decomposition for the non-Abelian gauge field \cite{Cho80,FN98,Shabanov99}: 
By introducing  a unit vector field $\bm{n}(x)$ with three components, i.e., 
$\bm{n}(x) \cdot \bm{n}(x) := n^A(x) n^A(x) = 1$ $(A=1,2,3)$,  
the non-Abelian gauge field $\mathscr{A}_\mu(x)$ in the SU(2) Yang-Mills theory is decomposed as
\begin{align}
 \mathscr{A}_\mu(x) 
= c_\mu(x) \bm{n}(x) +  g^{-1}  \partial_\mu \bm{n}(x)  \times \bm{n}(x)  + \mathbb{X}_\mu (x) . 
\end{align}
In what follows, we use the notation: 
$\mathbb{C}_\mu(x):=c_\mu(x) \bm{n}(x)$,  
$\mathbb{B}_\mu(x):=g^{-1} \partial_\mu \bm{n}(x) \times \bm{n}(x)$
and
$\mathbb{V}_\mu(x):=\mathbb{C}_\mu(x)+\mathbb{B}_\mu(x)$. 
By definition,  
$\mathbb{C}_\mu(x)$ is parallel to $\bm{n}(x)$, while $\mathbb{B}_\mu(x)$ is orthogonal to $\bm{n}(x)$.  We require $\mathbb{X}_\mu(x)$ to be orthogonal to $\bm{n}(x)$,  i.e., 
$\bm{n}(x) \cdot \mathbb{X}_\mu(x)=0$.  
We call $\mathbb{C}_\mu(x)$ the restricted potential, while $\mathbb{X}_\mu(x)$ is called the gauge-covariant potential and 
 $\mathbb{B}_\mu(x)$ is called the non-Abelian magnetic potential. 
In the naive Abelian projection,  
$\mathbb{C}_\mu(x)$ corresponds to the diagonal component, 
while $\mathbb{X}_\mu(x)$ corresponds to the off-diagonal component, 
apart from the vanishing magnetic part $\mathbb{B}_\mu(x)$.

Accordingly, the non-Abelian field strength $\mathscr{F}_{\mu\nu}(x)$ is decomposed as
\begin{align}
  \mathscr{F}_{\mu\nu} := \partial_\mu \mathscr{A}_\nu - \partial_\nu \mathscr{A}_\mu + g \mathscr{A}_\mu \times \mathscr{A}_\nu
= \mathbb{E}_{\mu\nu} + \mathbb{H}_{\mu\nu} + \hat{D}_\mu \mathbb{X}_\nu - \hat{D}_\nu \mathbb{X}_\mu + g \mathbb{X}_\mu \times \mathbb{X}_\nu , 
\end{align}
where we have introduced the covariant derivative in the background field $\mathbb{V}_\mu$ by 
$
  \hat{D}_\mu[\mathbb{V}]  \equiv \hat{D}_\mu  := \partial_\mu   + g \mathbb{V}_\mu \times  ,
$
and defined the two kinds of field strength:
\begin{align}
  \mathbb{E}_{\mu\nu} =& E_{\mu\nu} \bm{n}, \quad
E_{\mu\nu} := \partial_\mu c_\nu - \partial_\nu c_\mu ,
\\
  \mathbb{H}_{\mu\nu}  
=& \partial_\mu \mathbb{B}_\nu - \partial_\nu \mathbb{B}_\mu + g \mathbb{B}_\mu \times \mathbb{B}_\nu
 .
\end{align}
Due to the special definition of $\mathbb{B}_\mu$,  the magnetic field strength  $\mathbb{H}_{\mu\nu}$ is rewritten as
\begin{align}
  \mathbb{H}_{\mu\nu}  
&=   - g \mathbb{B}_\mu \times \mathbb{B}_\nu 
=  - g^{-1}   (\partial_\mu \bm{n} \times \partial_\nu \bm{n})  
=  H_{\mu\nu} \bm{n}, 
\\
H_{\mu\nu} &:=  - g^{-1} \bm{n} \cdot (\partial_\mu \bm{n} \times \partial_\nu \bm{n})  ,
\end{align}
where we have used a fact that $\mathbb{H}_{\mu\nu}$ is parallel to $\bm{n}$. 

\section{CFN decomposition on a lattice}

We define the CFN-Yang--Mills theory as the Yang-Mills theory written in terms of the CFN variables \cite{Cho80,FN98,Shabanov99}.
It has been shown \cite{KMS05} that the SU(2) CFN-Yang-Mills theory has the enlarged local gauge symmetry $\tilde{G}^{\omega,\theta}_{local}=SU(2)_{local}^{\omega} \times [SU(2)/U(1)]_{local}^{\theta}$ larger than the local gauge symmetry $SU(2)_{local}^{\omega}:=SU(2)_{local}^{I}$ in the original Yang--Mills theory. This is because we can rotate the CFN variable $\bm{n}(x)$ by angle $\bm{\theta}^{\perp}(x)$ independently of the gauge transformation of  $\mathscr{A}_\mu(x)$ by the parameter $\bm{\omega}(x)$.  
  In order to fix the whole enlarged local gauge symmetry $\tilde{G}^{\omega,\theta}_{local}$, we must impose sufficient number of gauge fixing conditions. 
Recently, it has been clarified \cite{KMS05} how the CFN-Yang--Mills theory can be equivalent to the original Yang-Mills theory by imposing a type of gauge fixing called the new Maximal Abelian gauge (nMAG) for fixing the extra local gauge invariance in the continuum formulation.

\subsection{New MAG and LLG}

Now we discuss how  the CFN decomposition  is implemented on a lattice by defining the unit color vector field $\bm{n}_{x}$ to generate the ensemble of $\bm{n}$-fields. 
In the whole of this paper, we restrict the gauge group to SU(2).

First of all, we generate the configurations of SU(2) link variables 
$\{ U_{x,\mu} \}$, 
\begin{align}
 U_{x,\mu}= \exp [ - ig\epsilon   \mathscr{A}_\mu(x) ] ,
\end{align}
using the standard Wilson action based on the heat bath method 
\cite{Creutz80} 
where $\epsilon$ is the lattice spacing and  $g$ is the coupling constant.  We use the continuum notation  for the Lie-algebra valued field variables, e.g., $\mathscr{A}_\mu(x)$, even on a lattice.

Next, we define the new Maximal Abelian gauge  (nMAG) on a lattice.  
By introducing a vector field ${\bm n}_{x}$ of a unit length with three components, we consider a functional $F_{nMAG}[U, {\bm n}; \Omega, \Theta]$  written in terms of  the gauge (link) variable $U_{x,\mu}$ and the color (site) variable $\bm{n}_{x}$ 
defined by 
\begin{align}
  F_{nMAG}[U, {\bm n}; \Omega, \Theta]
  := \sum_{x,\mu}  {\rm tr}({\bf 1}-{}^{\Theta}\bm{n}_{x} {}^{\Omega}U_{x,\mu} {}^{\Theta}\bm{n}_{x+\mu} {}^{\Omega}U_{x,\mu}^\dagger ) .
\end{align} 
Here we have introduced the enlarged gauge transformation: 
$
 {}^\Omega{}U_{x,\mu} := \Omega_{x} U_{x,\mu} \Omega_{x+\mu}^\dagger
$
for the  link variable 
$U_{x,\mu}$   
and
$
 {}^{\Theta}\bm{n}_{x} := \Theta_{x} \bm{n}_{x}^{(0)} \Theta_{x}^\dagger  
$ 
for an initial site variable $\bm{n}_{x}^{(0)}$
where gauge group elements $\Omega_{x}$ and $\Theta_{x}$   are independent SU(2) matrices on a site $x$.
The former corresponds to the  $SU(2)^{\omega}$ gauge transformation $(\mathscr{A}_\mu)^{\omega}(x)$  of the original potential: 
$
(\mathscr{A}_\mu)^{\omega}(x)=\Omega(x)[\mathscr{A}_\mu(x)+ig^{-1} \partial_\mu]\Omega^\dagger(x)
=\mathscr{A}_\mu(x)+D_\mu[\mathscr{A}]\bm{\omega}(x) + O(\bm{\omega}^2)$ for the 
$\Omega_{x}=e^{i g\bm{\omega}(x)}$, 
while 
the infinitesimal form of the latter reads   
$
{\bm n}^{\theta}(x)
  ={\bm n}(x)+g{\bm n}(x) \times {\bm\theta}(x) 
  ={\bm n}(x)+g{\bm n}(x) \times {\bm\theta}_\perp(x) 
$
for the adjoint $[SU(2)/U(1)]^{\theta}$ rotation
$\Theta_{x}=e^{i g\bm{\theta}(x)}$.

\begin{figure}[htbp]
\begin{center}
\includegraphics[height=4.5cm]{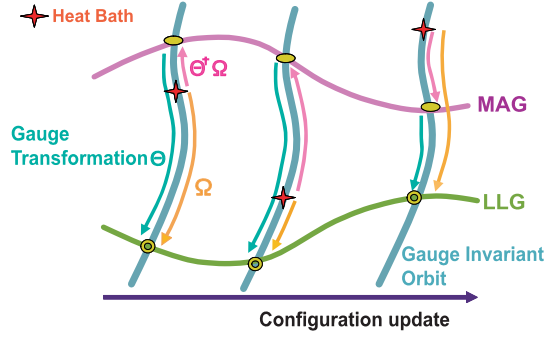}
\caption{\small 
Lattice CFN decomposition obtained by imposing nMAG and LLG. 
}
\label{fig:GF-orbit}
\end{center}
\end{figure}

After imposing the nMAG, the theory still has the local gauge symmetry $SU(2)_{local}^{\omega=\theta}:=SU(2)_{local}^{II}$, since the 'diagonal' gauge transformation ${\bm \omega}={\bm \theta}$ does not change the vaue of the functional $F_{nMAG}[U, {\bm n}; \Omega, \Theta]$. Therefore, ${\bm n}_{x}$ configuration can not be determined at this stage. 
In order to completely fix the gauge and to determine ${\bm n}_{x}$, we need to impose another gauge fixing condition for fixing  $SU(2)_{local}^{II}$.  
In this paper we choose the  conventional Lorentz-Landau gauge or Lattice Landau gauge (LLG) for this purpose. 
The LLG can be imposed by minimizing the function 
$F_{LLG}[U; \Omega]$:  
\begin{align}
  F_{LLG}[U; \Omega] = \sum_{x,\mu} {\rm tr}({\bf 1}-{}^{\Omega}U_{x,\mu}) 
  \rightarrow 1/4 \int d^4x \  [(\mathscr{A}_\mu)^{\omega}(x)]^2  
  \quad (\epsilon \rightarrow 0) ,
\end{align}
with respect to the gauge transformation $\Omega_{x}$ for the given link configurations $\{ U_{x,\mu} \}$, 
\begin{align}
  \min_{\Omega} F_{LLG}[U; \Omega] 
  \rightarrow 
  \min_{\omega} \int d^4x \  [(\mathscr{A}_\mu)^{\omega}(x)]^2  .
\end{align}
See Fig.~\ref{fig:GF-orbit}. 
In the continuum formulation, this is equivalent to imposing the gauge fixing condition
$\partial_\mu \mathscr{A}_\mu(x)=0$. 
The LLG fixes the local gauge symmetry $SU(2)_{local}^{\omega=\theta}=SU(2)^{II}_{local}$, while the LLG leaves the global symmetry SU(2)$_{global}^{\omega}=SU(2)^{II}_{global}$ intact.

Once  the link variable is identified with the CFN version:
\begin{align}
  U_{x,\mu} = \exp \{ - i \epsilon g[\mathbb{C}_\mu(x) + \mathbb{B}_\mu(x) + \mathbb{X}_\mu(x) ]\}  ,
  \label{LCFN}
\end{align}
it is straightforward to show that the functional $F_{nMAG}[U, {\bm n}; \Omega, \Theta]$ reduces in the naive continuum limit $\epsilon \rightarrow 0$ to the functional of the nMAG in the continuum formulation given in \cite{KMS05}:
\begin{align}
  \min_{\Omega, \Theta} F_{nMAG}[U, {\bm n}; \Omega, \Theta]
  \rightarrow \min_{\omega, \theta} \int d^4x \  [(\mathbb{X}_\mu)^{\omega, \theta}(x)]^2  .
\end{align} 
Therefore, we define the {\it lattice nMAG} by  minimizing the functional 
$F_{nMAG}[U, {\bm n}; \Omega, \Theta]$ with respect to the enlarged gauge transformation $\{ \Omega_{x} \}$ and $\{ \Theta_{x} \}$.
We call (\ref{LCFN}) the {\it lattice CFN decomposition}.

Then a remaining issue to be clarified is how to construct the ensemble of the color $\bm{n}$-fields used in defining $F_{nMAG}[U, {\bm n}; \Omega, \Theta]$ and $U_{x,\mu}$.  In what follows, we show that the desired color vector field $\bm{n}_{x}$ is constructed  from the interpolating gauge transformation matrix $\Theta_{x}$ by choosing the initial value  
${\bm n}_{x}^{(0)}=\sigma_3$ and 
\begin{align}
  \bm{n}_{x} := \Theta_{x} \sigma_3 \Theta_{x}^\dagger
  = n_{x}^A \sigma^A  , 
  \quad   n_{x}^A =  {\rm tr}[\sigma_A \Theta_{x} \sigma_3 \Theta_{x}^\dagger]/2   \quad  (A=1,2,3).
\end{align} 
A basic observation is that the functional $F_{nMAG}[U, {\bm n}; \Omega, \Theta]$ has another equivalent form $F_{MAG}[U; G]=F_{nMAG}[U, {\bm n}; \Omega, \Theta]$: 
\begin{align}
   F_{MAG}[U; G] 
  := \sum_{x,\mu}  {\rm tr}({\bf 1}-\sigma_3 \ {}^{G}{}U_{x,\mu} \ \sigma_3 \ {}^{G}{}U_{x,\mu}^\dagger ) 
    \rightarrow 
   \int d^4x \  [(A^a_\mu)^{G}(x)]^2  ,
\end{align}
with the identification
$G_{x}:=\Theta_{x}^\dagger \Omega_{x}$.
The minimization of $F_{MAG}[U; G]$ with respect to the gauge transformation $G_{x}=e^{ig\alpha_x}$ leads to the same form as the conventional MAG in the naive continuum limit:
\begin{align}
  \min_{G} F_{MAG}[U; G] 
    \rightarrow 
  \min_{\alpha} \int d^4x \  [(A^a_\mu)^{\alpha}(x)]^2  ,
\end{align}
if  the Cartan decomposition for $\mathscr{A}_\mu(x)$ has been used under the identification:
\begin{align}
  U_{x,\mu} = \exp \{ - i \epsilon g \mathscr{A}_\mu(x) \} 
= \exp \{ - i \epsilon g[a_\mu(x) T^3 + A_\mu^a(x) T^a ] \} ,
\end{align}
where $a_\mu$ and $A_\mu^a (a=1,2)$  are  respectively called  diagonal and  off-diagonal gauge fields with SU(2) generators $T^A=\sigma^A/2 (A=1,2,3)$. This procedure determines the configurations $\{ G_{x}^{*}\}$  achieving the minimum of $F_{MAG}[U; G]$. 
See Fig.~\ref{fig:GF-orbit}. 

On a gauge orbit, two representatives  on the two gauge-fixing hypersurfaces (MAG and LLG) are connected by the gauge transformation $\Theta_{x} \equiv \Omega_{x}^{*} (G_{x}^{*})^\dagger$.  Thus we can determine  a set of interpolating gauge-rotation matrices $\{ \Theta_x \}$ to construct ${\bm n}_{x}$ ensemble. 
In fact, the color vector $\bm{n}_{x}$ constructed in this way represents 
a real-valued vector $\vec{n}_{x}=(n_{x}^1,n_{x}^2,n_{x}^3)$ of  unit length  with three components, and transforms in the adjoint representation under the gauge transformation II.
By imposing simultaneously the nMAG and the LLG  in this way, we can completely fix the local gauge invariance 
$\tilde{G}^{\omega,\theta}_{local}$
of the lattice CFN-Yang--Mills theory. 
It is remarkable that, even after the complete gauge fixing, the global gauge (color) symmetry $SU(2)_{global}^{\omega=\theta}$ is unbroken.

\subsection{Numerical simulations}

Our numerical simulations are performed as follows. 
In the continuum formulation,  the CFN variables were introduced as a change of variables in such a way that they do not break the global gauge symmetry  $SU(2)_{global}^{II}$  or "color symmetry", corresponding to the global  gauge symmetry $SU(2)_{global}^{I}$ in the original Yang-Mills theory. 
Hence the nMAG can be imposed in terms of the CFN variables without breaking the color symmetry. 
This is a crucial difference between the nMAG based on the CFN decomposition and the conventional MAG based on the ordinary Cartan decomposition which breaks the  $SU(2)_{global}$ explicitly. 
Therefore, we must perform the numerical simulations so as to preserve the color symmetry as much as possible.%
\footnote{ 
Whether the color symmetry is spontaneously broken or not is another issue to be investigated separately.  Our construction allows us to study this issue without breaking it explicitly. 
}
This is in fact possible as follows.

Remember that the nMAG on a lattice is achieved by repeatedly performing the gauge transformations. 
In order to preserve the global SU(2) symmetry,  
we adopt a random gauge transformation only in the first sweep among the whole sweeps of gauge transformations in the standard iterative gauge fixing procedure for the LLG.
This procedure moves an ensemble of unit vectors $\bm{n}_{x}$ to a random ensemble of $\bm{n}_{x}$ which is far away from $\bm{n}_{x}=(0,0,1)$. 
Then we search for the local minima around this configuration of $\bm{n}_{x}$ by performing the successive gauge transformations. 
The first random gauge transformation as well as the subsequent gauge transformations are accumulated to obtain the gauge transformation matrix $G$ by which $\bm{n}$ is constructed.

Beginning with the  MAG and ending with the LLG in this way, we can impose both nMAG and LLG   simultaneously.

Our numerical simulations are performed on the lattice  with the   
lattice size $L^4=8^{4}$ and $16^{4}$
by using the standard Wilson action  for the gauge coupling $\beta=2.2\sim 2.45$ and periodic boundary conditions. 
Staring with cold initial condition and thermalizing 30*100 (resp. 50*100)
sweeps, we have obtained  50 (resp. 200) configurations (samples) for $8^4 $(resp. $16^4$) lattice at intervals of 100 sweeps.
For LLG and MAG, we have used the over relaxation algorithm.

\begin{table}[h]
\caption{The magnetization $<n^A_x>$ on the $16^4$ lattice at $\beta=2.4$.}
\label{magnet-lma6}
\begin{center}
\begin{tabular}{cll}\hline
      & Mean value & Jack knife error(JKbin=2) \\ \hline
$<n^1>$ & -0.0069695  & $\pm$ 0.010294   \\
$<n^2>$ & 0.011511 & $\pm$ 0.015366 \\
$<n^3>$ & 0.0014141 & $\pm$ 0.013791 \\ \hline
\end{tabular}
\label{table:vevn}
\end{center}
\end{table}

The data of numerical simulations in Table~\ref{table:vevn} show the vanishing vacuum expectation value  
\begin{align}
  \left< n^A_x \right> = 0  \quad (A=1,2,3) . 
\end{align}

\begin{figure}[htbp]
\begin{center}
\includegraphics[height=5cm]{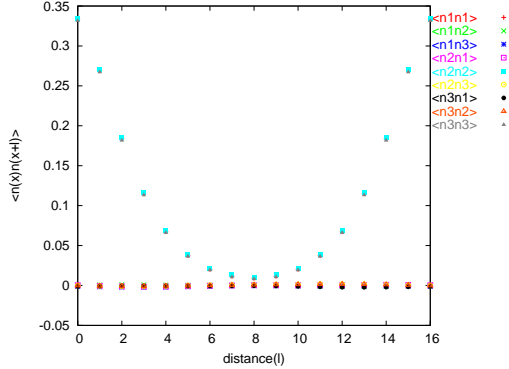}
\caption{\small 
The plots of two-point correlation functions 
$\left< n_x^A n_0^B \right>$ for $A,B=1,2,3$
along the lattice axis on the $16^4$ lattice at $\beta=2.4$.
}
\label{fig:2ptf}
\end{center}
\end{figure}

Moreover, we have measured the two-point correlation functions defined by $\left< n^A_x n^B_0 \right>$, see Fig.~\ref{fig:2ptf}.
The two-point correlation functions $\left< n^A_x n^A_0 \right>$  (no summation over $A$) exhibit almost the same behavior in all the directions ($A=1,2,3$), while $\left< n^A_x n^B_0 \right>$ ($A\not=B$) vanish. Thus, 
we have obtained the correlation function
$\left< n_x^A n_0^B \right>=\delta^{AB}D(x)$ respecting color symmetry.

These results indicate that {\it the global $SU(2)$ symmetry (color symmetry) is   unbroken in our main simulations}.
This is the first remarkable result.

\subsection{Discriminating our approach from the others}

According to our viewpoint \cite{KMS05}, the $\bm{n}_{x}$ field  must respect the global $SU(2)^{II}$ symmetry. If this is not the case, the field $\bm{n}_{x}$ can not be identified with the color vector field in the CFN decomposition of the original gauge potential from our viewpoint \cite{KMS05}. This is a crucial point. 
This is  in sharp contrast to the previous approaches \cite{DHW02,Shabanov01}.  
Although the similar technique of constructing the unit vector field $\bm{n}_x$ from a $SU(2)$ matrix $G_{x}$ has already  appeared, e.g., in \cite{DHW02,Shabanov01,IS00}, there is a crucial conceptual difference between our approach and others.  

We can perform the global SU(2) rotation at will, since it is not prohibited in our setting.
However, the previous numerical simulations are performed only in a restricted setting where LLG and MAG are close to each other \cite{DHW02} by imposing LLG as a preconditioning,
in the sense that the matrices $G$ connecting LLG and MAG are on average close to the unit ones.
That is to say, 
$G_x^A \cong 0$ $(A=1,2,3)$, i.e., $G_x \cong g_x^0 I$, 
for the parameterization of $SU(2)$ matrices, 
$
  G_x = G_x^0 I + i G_x^A \sigma^A
$, \quad
$  
  G_x^0, G_x^A \in \mathbb{R}
$, \quad
$
   \sum_{\mu=0}^{3} (G_x^\mu)^2 = 1 .
$
Then it has been observed that 
$\bm{n}_x \cong \sigma^3$ or $n_x^A \cong (0,0,1)$, namely,  
 $\bm{n}_x$ are  aligned in the positive 3-direction and hence the non-vanishing vacuum expectation value is observed as 
\begin{align}
  \left< n^A_x \right> = M \delta^{A3} . 
\end{align}
This implies that the global SU(2) symmetry is broken explicitly to a global U(1),
 $SU(2)_{\text{global}} \rightarrow U(1)_{\text{global}}$. 
In the two-point correlation functions, the exponential decay has been observed for 
the parallel propagator  
\begin{align}
  \left< n_x^3 n_0^3 \right> \sim \left< n_0^3 \right>\left< n_0^3 \right> + c e^{-m|x|} = M^2 + c e^{-m|x|} , 
\end{align}
and for the perpendicular propagator
\begin{align}
   \sum_{a=1}^{2} \left< n_x^a n_0^a \right> \sim  c' e^{-m'|x|} ,
\end{align}
with $m$ and $m'$ being different to each other. 
This result was reported by \cite{DHW02} ( and confirmed also by our preliminary simulations \cite{DSB04}).

In our approach we can identify the lattice field $\bm{n}_x$ as a lattice version of the CFN field variable $\bm{n}(x)$ obtained by the CFN decomposition of the gauge potential $\mathscr{A}_\mu(x)$ in the original Yang-Mills theory in agreement with the new viewpoint \cite{KMS05}. Moreover, we do not assume any effective theory of Yang-Mills theory written in terms of the unit vector field $\bm{n}_x$, such as the Skyrme--Faddeev model.

\section{Monopole current on a lattice}

\subsection{Gauge-invariant monopole current on the lattice}

We can define the {\it gauge-invariant} monopole current  by
\begin{align}
  k^\mu(x) :=& \partial_\nu {}^*G^{\mu\nu}(x) 
=   (1/2) \epsilon^{\mu\nu\rho\sigma}\partial_{\nu}
              G_{\rho\sigma}(x) ,
\nonumber\\
 G_{\mu\nu} :=& \bm{n} \cdot (\mathbb{E}_{\mu\nu}+\mathbb{H}_{\mu\nu})
=  E_{\mu\nu}+H_{\mu\nu} 
= \partial_\mu c_\nu - \partial_\nu c_\mu  -  g^{-1}  \bm{n} \cdot  (\partial_\mu \bm{n} \times \partial_\nu \bm{n}) .
\label{latCFN-monop}
\end{align}
It is remarkable that the CFN monopole current defined in this way is gauge invariant. 
This is because $G^{\mu\nu}(x)$ is invariant under the gauge transformation II, see  \cite{KMS05}. 
This is not the case for the respective quantity, $E_{\mu\nu}(x)$ and $H_{\mu\nu}(x)$.

We define a lattice version $K_{\mu}({}^*n)$ of the dual 1-form $k^\mu$  on the dual lattice \cite{Sijs91} by
\begin{eqnarray}
g^{-1} K_{\mu}({}^*n) &\equiv& 
-(1/2)\epsilon_{\mu\nu\rho\sigma}\partial_{\nu}
              G_{\rho\sigma}(n+\hat{\mu}) ,
\label{latCFN-monop-2}
\end{eqnarray}
where ${}^*n$ is a site on the dual lattice ${}^*n=n+(\frac{1}{2}\epsilon,\frac{1}{2}\epsilon,\frac{1}{2}\epsilon,\frac{1}{2}\epsilon)$ for a site $n$ on an original lattice 
and the derivative $\partial_{\nu}$ on a lattice is defined  by the forward lattice derivative:
$ 
\partial_{\nu} G_{\rho\sigma}(n)  \equiv  
G_{\rho\sigma}(n+\hat{\nu})-G_{\rho\sigma}(n)
$. 
Here the gauge coupling constant $g$ was extracted from the definition. 
The explicit form of $K_1({}^*n)$ reads, e.g., 
\begin{eqnarray}
g^{-1} K_{1}({}^*n)&=&
\partial_{2} G_{34}(n+\hat{1})
-\partial_{3} G_{24}(n+\hat{1})
+\partial_{4} G_{23}(n+\hat{1}) \\
&=& G_{34}(n+\hat{1}+\hat{2})-G_{34}(n+\hat{1})\nonumber\\
&& -G_{24}(n+\hat{1}+\hat{3})+G_{24}(n+\hat{1})\nonumber\\
&& +G_{23}(n+\hat{1}+\hat{4})-G_{23}(n+\hat{1}) .
\label{latCFN-monop-4}
\end{eqnarray}

\subsection{Simulation results: comparison with DT monopole}

\begin{figure}[htbp]
\begin{center}
\includegraphics[height=5cm]{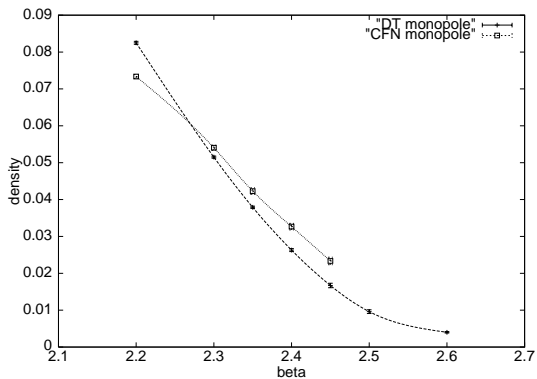}
\caption{\small 
Comparison  of CFN and DT magnetic-current densities versus $\beta$. 
The CFN density is plotted for $\rho_{mon}' := \rho_{mon}/2\pi$ to be consistent with the Dirac quantization condition.}
\label{fig:monopole-density}
\end{center}
\end{figure}

The construction of the monopole current (\ref{latCFN-monop}) on a lattice is performed according to the standard method \cite{Sijs91}. 
This CFN monopole should be compared with the conventional DT definition  \cite{DT80} of the magnetic monopole  which has been extensively used in the previous studies of QCD monopole.
As a preliminary trial, we have measured the magnetic-current density $\rho_{mon}$ defined by
\begin{align}
  \rho_{mon} :=  \left< \sum_{x,\mu} |k_\mu(x)| \right>/4V,
\end{align}
where $|k_\mu(x)|$ is the absolute value of $k_\mu(x)$ and $V$ is the volume of the lattice, i.e., the total number of   lattice sites $V=L^4$.  We call $\rho_{mon}$ the CFN magnetic-charge density. 
More extensive studies including other physical quantities  will be given in a subsequent paper. 
In Fig.~\ref{fig:monopole-density}, we have plotted both the CFN magnetic-charge density $\rho_{mon}$ and the DT one for the value of $\beta = 2.2 \sim 2.6$. 
Fig.~\ref{fig:monopole-density} shows that the magnitude of the CFN magnetic-current density is nearly equal to the DT one, and that both densities exhibit the same $\beta$ dependence. 

\subsection{Contribution from the electric field}

Finally, we consider the anatomy of the CFN magnetic-current density $\rho_{mon}$ which includes the 'electric' part $\rho_{E}$ and 'magnetic' one $\rho_{H}$. \footnote{
The distinction between electric and magnetic is merely our convention. 
The magnetic-current density is different from the magnetic-charge density whose integration over the three volume gives the magnetic charge $Q_m =\int d^3x k_0$. 
Note that $\rho_{mon}:=\rho_{G}=\rho_{E+H}\not=\rho_{E}+\rho_{H}$ where $\rho_{E}$ and $\rho_{H}$ is defined by replacing $G_{\mu\nu}$ with $E_{\mu\nu}$ and $H_{\mu\nu}$ respectively. This forms a striking contrast to $Q_m =Q_m^E + Q_m^H $, which comes from  
$k_\mu=k_\mu^E+k_\mu^H$.
}
To estimate the electric part, we must specify the contribution from the restricted potential 
$c_\mu(x) := \bm{n}(x) \cdot \mathscr{A}_\mu(x)$ 
to the magnetic-current density. 
For this purpose, $c_\mu(x)$ is extracted as follows. 
From the definition  of the link variable 
 $U_{x,\mu}=e^{-i{\bf A}_{\mu}(x)}$ (${\bf A}_{\mu}(x)\equiv ag A_{\mu}(x)$), 
Yang-Mills field ${\bf A}_{\mu}(x)$ is given up to the higher order corrections in the lattice spacing $a$ by
\begin{equation}
{\bf A}_{\mu}(x) = i(U_{x,\mu}-U^{\dagger}_{x,\mu})/2 .
\label{lat-full-CFN-monop-3}
\end{equation}
Substituting the parameterization of $U_{x,\mu}$:   
\begin{equation}
 U_{x,\mu}=U^0_{x,\mu}1+iU^A_{x,\mu}\sigma^A  \quad (A=1,2,3)
\end{equation}
into 
(\ref{lat-full-CFN-monop-3}), we obtain
\begin{align}
{\bf A}_{\mu}(x)={\bf A}^A_{\mu}(x) \sigma^A/2 , \quad 
{\bf A}^A_{\mu}(x) = -2 U^A_{x,\mu} .
\label{lat-full-CFN-monop-4}
\end{align}
Once the configurations 
 $\{n^1_x,n^2_x,n^3_x\}$ and 
$\{U^0_{x,\mu},U^1_{x,\mu},U^2_{x,\mu},U^3_{x,\mu}\}$ 
of $n$ and $U$ are generated by the simulations, 
the configuration $\{c_{\mu}(x)\}$ of the restricted potential  
$
c_{\mu}(x) = n^A(x){\bf A}^A_{\mu}(x) = Tr[{\bf n}(x){\bf A}_{\mu}(x)]
$
is obtained from the expression:
\begin{equation}
c_{\mu}(x) = -2\{ n^1_xU^1_{x,\mu} + n^2_xU^2_{x,\mu} + n^3_xU^3_{x,\mu} \} .
\end{equation}
The electric field strength  $E_{\mu\nu}(x)$ is defined from the restricted potential $c_{\mu}(x)$ by 
$
E_{\mu\nu}(x)= {\partial}_{\mu}c_{\nu}(x)-{\partial}_{\nu}c_{\mu}(x)
\label{lat-full-CFN-monop-7}
$
which is implemented on a lattice by replacing the derivative with the naive finite difference on a lattice:
$
E_{\mu\nu}(x)= c_{\nu}(x+\hat{\mu}) - c_{\nu}(x) - c_{\mu}(x+\hat{\nu}) + c_{\mu}(x)
$.

At $\beta=2.35$,  
the total monopole density $\rho_{G}=\rho_{E+H}$ has the value:  
\begin{align}
\rho_{mon} \equiv \rho_{G} = \rho_{E+H} = 0.2694 \pm 0.0057 ,
\label{lat-full-CFN-monop-11}
\end{align}
while
the monopole density $\rho_{H}$ calculated only from the $H_{\mu\nu}(x)$ is
\begin{align}
\rho_{H} = 0.2694 \pm 0.0057 ,
\label{lat-full-CFN-monop-10}
\end{align}
which can be calculated from the configuration of the color field $\bm{n}_{x}$ alone.
Therefore,  
the (gauge-invariant)  magnetic-current density $\rho_{G}=\rho_{E+H}$
is dominated by $\rho_{H}$  which is not gauge invariant. 
In other words, the electric part $E_{\mu\nu}(x)$ does not contribute to the magnetic-current  in our setting. 

The configuration of the monopole current located at the origin $0$ of the lattice in the first sweep is given by
\begin{align}
(k_1,k_2,k_3,k_4)
=(-1.1456, 0.7506, 2.8916, -1.1075).
\label{lat-full-CFN-monop-12}
\end{align}
Here, the contribution from the electric field strength $E_{\mu\nu}(x)$ is negligibly small:
\begin{align}
 (k^E_1,k^E_2,k^E_3,k^E_4)
=(-3.33\times 10^{-16}, 
-5.55\times 10^{-17}, -1.11\times 10^{-16}, 1.11\times 10^{-16}) ,
\label{lat-full-CFN-monop-13}
\end{align}
whereas the contribution from the magnetic field strength $H_{\mu\nu}(x)$ leads to 
\begin{align}
(k^H_1,k^H_2,k^H_3,k^H_4)=(-1.1456, 0.7506, 2.8916, -1.1075).
\label{lat-full-CFN-monop-14}
\end{align}
Thus, we conclude that the contribution of the electric field strength $E_{\mu\nu}$ to the magnetic current $k_\mu$ is negligible 
and the magnetic monopole is provided only from the magnetic field strength  $H_{\mu\nu}$ in our setting, as expected in \cite{KKMSS05}. 
This implies that the color vector $\bm{n}(x)$ is not aligned in the same direction in the color space so that it fails to produce the non-zero magnetic current $k_\mu$ through $E_{\mu\nu}$. Rather, the color vector $\bm{n}(x)$  represents the hedgehog like configuration to produce the whole magnetic current through $H_{\mu\nu}$. 
The configuration of $\bm{n}$ obtained in our simulations does not lead to the singularities in  the electric part $E_{\mu\nu}$ and that the electric part $E_{\mu\nu}$ does not contribute to the monopole current.
However, this does not necessarily mean that the 
restricted potential $c_{\mu}(x)$ itself is small. 
In fact, $c_\mu(x)$ is not small at the origin:
\begin{align}
(c_1,c_2,c_3,c_4)=( -0.4757, -0.2416, 0.3047, 0.0348).
\label{lat-full-CFN-monop-15}
\end{align}
The above results can be understood if we recall the 't Hooft--Polyakov magnetic monopole in Georgi--Glashow model under the identification of the unit Higgs scalar field $\hat\phi(x)$ with the color vector field $\bm{n}(x)$.


\section{Conclusion and Discussion}

In this paper, we have shown how to implement the CFN decomposition (change of variables) of SU(2) Yang--Mills theory on a lattice, according to a new viewpoint proposed in \cite{KMS05}.  A remarkable point is that our approach can preserve both the local SU(2) gauge symmetry and the color symmetry (global SU(2) symmetry) even after imposing a new type of gauge fixing (called the nMAG) which is regarded as a constraint  to reproduce the original Yang-Mills theory.  
Moreover, we have succeeded to perform the numerical simulations in such a way that the color symmetry is unbroken. 

We have given a new definition of the magnetic monopole-current called the CFN monopole current by using the CFN variables on a lattice just constructed. 
An advantage of our definition of magnetic monopole is that it is  SU(2) gauge invariant. 
Our numerical simulations show that the CFN monopole gives nearly the same value as given by DT monopole, suggesting the equivalence of two definitions on a lattice. 
Therefore, the CFN monopole is expected to be used, instead of DT monopole, to study the confinement phenomena of SU(2) QCD and reproduce the monopole dominance in the string tension, the monopole action as a low-energy effective action of QCD, finite-temperature phase transition (confinement-deconfinement transition).

Our construction of the magnetic current is more similar to the original   Abelian projection of 't Hooft rather than that of DT. 
However, the construction of the CFN  monopole current given in this paper does not guarantee the integer-valued magnetic charge realized by DT monopole. 
The magnetic charge is indeed the real valued, in contrast with the conventional DT monopole. 
This disadvantage can be cured by converting it to the compact variable so as to guarantee the quantized magnetic charge from the beginning, as given in a subsequent paper.  It will be interesting to examine the contribution of the CFN monopole to the Wilson loop for confirming the monopole dominance in confinement.

\section*{Acknowledgments}
The numerical simulations have been done on a supercomputer (NEC SX-5) at  Research Center for Nuclear Physics (RCNP), Osaka University.
This project is also supported in part by the Large Scale Simulation Program No.133 (FY2005) of High Energy Accelerator Research Organization (KEK). 
K.-I. K. is financially supported by 
Grant-in-Aid for Scientific Research (C)14540243 from Japan Society for the Promotion of Science (JSPS), 
and in part by Grant-in-Aid for Scientific Research on Priority Areas (B)13135203 from
the Ministry of Education, Culture, Sports, Science and Technology (MEXT).




\end{document}